\documentclass[12pt]{article}
\usepackage{newinutile3}
\usepackage{amsmath,amssymb}
\usepackage{mathrsfs}
\usepackage{epsfig,graphics,color,calc}
\usepackage{cite}
\usepackage{rotating}

\newtheorem{theorem}{Theorem}

\newtheorem{lemma}[theorem]{Lemma}

\newcommand{\newatop}[2]{\genfrac{}{}{0pt}{}{#1}{#2}}
\newcommand{\qed}{$\phantom.$\hfill $\Box$\bigskip}
\newcommand{\rr}[1]{{\normalfont\textrm{#1}}}
\newcommand{\cc}[1]{{\mathcal{#1}}}
\newcommand{\bb}[1]{{\mathbb{#1}}}

\newcommand{\stab}{{x_0}}

\newcommand{\puno}{{\normalfont\textbf{u}}}
\newcommand{\muno}{{\normalfont\textbf{d}}}

\newlength{\pecettawidth}
\def\thesubsection{{\normalsize
{\arabic{section}.\arabic{subsection}}}}

\renewcommand{\tilde}{\widetilde}

\renewcommand\refname{{\normalsize\textbf{References}}}

\newcommand{\eC}{\mathbf{c}^\textrm{e}}
\newcommand{\oC}{\mathbf{c}^\textrm{o}}
\newcommand{\cC}{\mathbf{c}}

\begin{document}

\title{Sum of exit times in series of metastable states in 
Probabilistic Cellular Automata}
\author{Emilio N.M.\ Cirillo}
\affiliation{Dipartimento di Scienze di Base e Applicate per
             l'Ingegneria, Sapienza Universit\`a di Roma,
             via A.\ Scarpa 16, I--00161, Roma, Italy.}
\email{emilio.cirillo@uniroma1.it}

\author{Francesca R.\ Nardi}
\affiliation{Department of Mathematics and Computer Science,
             Eindhoven University of Technology,
             P.O.\ Box 513, 5600 MB Eindhoven, The Netherlands.}
\affiliation{Eurandom, P.O.\ Box 513, 5600 MB, Eindhoven, The Netherlands.}
\email{F.R.Nardi@tue.nl}

\author{Cristian Spitoni}
\affiliation{Department of Mathematics,
              Budapestlaan 6,
                  3584 CD Utrecht,The Netherlands}
\email{C.Spitoni@uu.nl}
\maketitle

\begin{abstract}
Reversible Probabilistic Cellular Automata are a special class 
of automata whose stationary behavior is described by Gibbs--like
measures. For those models the dynamics can be trapped for a very 
long time in states which are very different from the ones typical 
of stationarity.
This phenomenon can be recasted in the framework of metastability 
theory which is typical of Statistical Mechanics. 
In this paper we consider a model presenting two not degenerate in 
energy
metastable states which form a series, in the sense that, 
when the dynamics is started at one of them, before reaching 
stationarity, the system must necessarily visit the second one.
We discuss a rule for combining the exit times
from each of the metastable states. 
\end{abstract}

\section{Introduction}
\label{intro}\\
Cellular Automata (CA) are discrete--time dynamical systems on a 
spatially extended discrete space, see, e.g., \cite{Kari} and 
references therein.
Probabilistic Cellular Automata (PCA) are CA straightforward generalizations 
where the updating rule is stochastic (see \cite{sirakoulis2012cellular,LMS,marcovici}).
Strong relations exist between PCA and the general equilibrium statistical 
mechanics framework \cite{GJH,LMS}. Traditionally, 
the interplay between disordered global states 
and ordered phases has been addressed, but,
more recently, it has been remarked that even from the 
non--equilibrium point of view analogies between statistical mechanics 
systems and PCA deserve attention \cite{CN}.

In this paper we shall consider a particular class of PCA, called 
\textit{reversible} PCA.
Here the word reversible is used in the sense
that the detailed balance condition is satisfied with respect to a
suitable Gibbs--like measure (see the precise definition given just below
equation \eqref{ham})
defined via a translation invariant 
multi--body potential.
Such a measure depends on a parameter 
which plays a role similar to that played by the temperature 
in the context of statistical mechanics systems and which, for such a 
reason, will be called temperature. 
In particular, 
for small values of such a temperature, the dynamics of the PCA 
tends to be frozen in the local minima of the Hamiltonian 
associated to the Gibbs--like measure. Moreover, in suitable low temperature regimes (see \cite{PY}) the transition probabilities of the PCA become very small and the effective change of  a cell's state becomes rare, so that the
PCA dynamics becomes almost  a sequential one.

It is natural to pose, even for reversible PCA's, the question 
of metastability which arose overbearingly
in the history of thermodynamics and statistical mechanics since the 
pioneering works due to van der Waals. 

Metastable states are observed when a 
physical system is close to a first order phase transition. Well--known
examples
are super--saturated vapor states and magnetic hysteresis~\cite{OV}.
Not completely rigorous approaches based on equilibrium states have been 
developed in different fashions.
However, a fully mathematically rigorous theory, has been obtained 
by approaching the problem from a dynamical point of view. 
For statistical mechanics systems on a lattice a dynamics is 
introduced (a Markov process having the Gibbs measure as 
stationary measure) and metastable states are interpreted as those 
states of the system such that the corresponding time needed to 
relax to equilibrium is the longest one on an exponential scale
controlled by the inverse of the temperature. 
The purely dynamical point of view revealed itself extremely powerful and 
led to a very elegant definition and characterization of 
the metastable states. The most important results in this
respect have been summed up in~\cite{OV}.

The dynamical description of metastable states suits perfectly for 
their generalization to PCA \cite{CN,CNS01,CNS02,CNS03}. 
Metastable states have been investigated 
for PCA's in the framework of the so called \textit{pathwise approach}
\cite{OS,OV,MNOS}.
It has been shown how it is possible to characterize the exit 
time from the metastable states up to an exponential precision and 
the typical trajectory followed by the system during its 
transition from the metastable to the stable state. 
Moreover, it has also been shown how to apply the so called 
\textit{potential theoretic approach} \cite{BEGK} to compute 
sharp estimates for the exit time \cite{NS} of a specific PCA.

More precisely, the exit time from the metastable state 
is essentially in the form $K\exp\{\Gamma/T\}$ where $T$ is 
the temperature, 
$\Gamma$ is the energy cost of the best (in terms 
of energy) 
paths connecting the metastable state to the stable one, and $K$ 
is a number which is inversely connected to the number of possible 
best paths that the system can follow to perform its transition 
from the metastable state to the stable one. 
Up to now, in the framework of PCA models, the constant $K$ has
been computed only in cases in which the metastable state is unique.
The aim of this work is to  consider a PCA model in which
two metastable states are present. Similar results in the
framework of the Blume--Capel model with Metropolis dynamics
have been proved in \cite{CNS04,Landim}.

We shall consider the PCA studied in \cite{CN} 
which is characterized by the presence of two metastable states. 
Moreover, starting from one of them, the system, in order to perform 
its transition to the stable state, must necessarily visit the 
second metastable state. The problem we pose and solve in this 
paper is that of studying how the exit times from the two metastable
states have to be combined to find the constant $K$ characterizing 
the transition from the first metastable state to the stable one. 
We prove that $K$ is the sum of the two constants associated 
with the exit times from the two metastable states.

The paper is organized as follows, 
in Section~\ref{s:mod} we define the class of models considered,
in Section~\ref{meta} we state  the main result and recall the main mathematical tools used in its proof, and 
in Section~\ref{s:proof} we sketch the proof of the main theorem of the paper.


\section{The model}
\label{s:mod}
\par
\begin{figure}[t]

\setlength{\unitlength}{1.5pt}
\hspace{3.5cm}
 \begin{picture}(-10,55)(-60,-5)
 \thinlines
 \multiput(-5,0)(0,10){5}{\put(0,0){\line(1,0){50}}}
 \multiput(0,-5)(10,0){5}{\put(0,0){\line(0,1){50}}}
 \put(20,30){\circle*{2}}
 \put(30,20){\circle*{2}}
 \put(20,10){\circle*{2}}
 \put(10,20){\circle*{2}}
 \put(21,20){${\scriptscriptstyle 0}$}
 \thicklines
 \put(5,15){\line(0,1){10}}
 \put(15,15){\line(0,1){10}}
 \put(25,15){\line(0,1){10}}
 \put(15,15){\line(1,0){10}}
 \put(5,25){\line(1,0){10}}
 \put(15,25){\line(0,1){10}}
 \put(15,25){\line(1,0){10}}
 \put(15,35){\line(1,0){10}}
 \put(25,35){\line(0,-1){10}}
 \put(25,25){\line(1,0){10}}
 \put(35,25){\line(0,-1){10}}
 \put(35,15){\line(-1,0){10}}
 \put(25,15){\line(0,-1){10}}
 \put(25,5){\line(-1,0){10}}
 \put(15,5){\line(0,1){10}}
 \put(15,15){\line(-1,0){10}}
\end{picture}
\caption{Diamond $V(0)$ for the nearest neighbor model.}
\label{fig:modelli}
\end{figure}
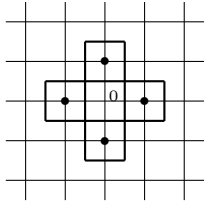

\noindent
In this section we introduce the basic notation and we define the model of reversible
PCA which will be studied in the sequel. Consider the 
two--dimensional torus $\Lambda=\{0,\dots,L-1\}^2$,
with $L$ even\footnote{The side length of the lattice is chosen even so that 
it will possible to consider configurations in which the plus and the minus 
spins for a checkerboard and fulfill the periodic boundary conditions.}, 
endowed with the Euclidean metric.
Associate a variable $\sigma(x)=\pm1$
with each site $x\in\Lambda$ and let $\mathcal{S}=\{-1,+1\}^{\Lambda}$ be the 
configuration space.
Let $\beta>0$ and $h\in(0,1)$.
Consider the Markov chain $\sigma_n$, with $n=0,1,\dots$,
on $\mathcal{S}$ with  transition matrix:
\begin{equation}
\label{markov}
p(\sigma,\eta)
=\prod_{x\in\Lambda}p_{x,\sigma}\left(\eta(x)\right)\;\;\;
\forall\sigma,\eta\in\mathcal{S}
\end{equation}
where, for $x\in\Lambda$ and $\sigma\in\mathcal{S}$,
$p_{x,\sigma}(\cdot)$ is the probability measure on $\{-1,+1\}$
defined as 
\begin{equation}
\label{rule}
p_{x,\sigma}(s)
=
\frac{1}{1+\exp\left\{-2\beta s(S_\sigma(x)+h)\right\}}
=
\frac{1}{2}
\left[1+s\tanh\beta \left(
S_\sigma(x)+h\right)\right]
\end{equation}
with $s\in\{-1,+1\}$ and
$S_{\sigma}(x)=\sum_{y\in\Lambda}K(x-y)\,\sigma(y)$
where 
$K(x-y)=1$  if $|x-y|=1$, and $K(x-y)=0$ otherwise.
The probability $p_{x,\sigma}(s)$ for the spin $\sigma(x)$ to be equal to $s$
depends only on the values of the spins of $\sigma$ 
on the diamond $V(x)$ centered at $x$, as shown  
in Fig.~\ref{fig:modelli} (i.e., the von Neumann neighborhood 
without the center). 

At each step of the dynamics all the spins of the system are updated 
simultaneously according to the probability distribution 
\eqref{rule}. This means the the value of the spin 
tends to align with the local field $S_\sigma(x)+h$: $S_\sigma(x)$
mimics a ferromagnetic interaction effect among spins, 
whereas $h$ is an external \emph{magnetic field}. 
Such a field, as said before, is chosen smaller than one otherwise 
its effect would be so strong to destroy the metastable
behavior. When $\beta$ is large the tendency to align with the 
local field is perfect, while for $\beta$ small also spin updates 
against the local filed can be observed with a not too small probability. 
Thus $\beta$ can be interpreted as the inverse of the \emph{temperature}. 

This kernel $K$ choice leads to the \textit{nearest neighbor PCA}
model studied in \cite{CN}.
The Markov chain $\sigma_n$ defined in (\ref{markov}) 
updates all the spins simultaneously and independently at any time
and it satisfies the 
\textit{detailed balance} property
$p(\sigma,\eta)\,e^{-\beta H(\sigma)}=
 p(\eta,\sigma)\,e^{-\beta H(\eta)}$
with
\begin{equation}
\label{ham}
H(\sigma)=
-h\sum_{x\in\Lambda}\sigma(x)
-\frac{1}{\beta}\sum_{x\in\Lambda}\log\cosh\left[\beta
\left(
S_{\sigma}(x)+h\right)\right]
\;\;.
\end{equation}
This is also expressed by saying that the dynamics 
is \textit{reversible} 
with respect to the Gibbs measure
$\mu(\sigma)=\exp\{-\beta H(\sigma)\}/Z$
with
$Z=\sum_{\eta\in\mathcal{S}}\exp\{-\beta H(\eta)\}$.
This property implies that 
$\mu$ is stationary, i.e., 
$\sum_{\sigma\in\mathcal{S}}\mu(\sigma)p(\sigma,\eta)=\mu(\eta)$.

It is important to remark that, 
although the dynamics is reversible, the probability 
$p(\sigma,\eta)$ cannot be expressed in terms of  
$H(\sigma)-H(\eta)$, as it usually happens for the serial Glauber dynamics,
typical of Statistical Mechanics. Thus,  
given $\sigma,\eta\in\mathcal{S}$, we define the \textit{energy cost} 
\begin{equation}
\label{defdelta}
\Delta(\sigma,\eta)=
 -\lim_{\beta\to\infty}\frac{\log p(\sigma,\eta)}{\beta}
 =
\!\!\!\!\!\!\!\!\!
\sum_{\newatop{x\in\Lambda:}{\eta(x)[S_\sigma(x)+h]<0}}
\!\!\!\!\!\!\!\!\!
2|S_\sigma(x)+h|
\end{equation}
Note that $\Delta(\sigma,\eta)\ge0$ and $\Delta(\sigma,\eta)$ is not
necessarily equal to $\Delta(\eta,\sigma)$;
it can be proven, see \cite[Sect.~2.6]{CNS01}, that 
\begin{equation}
\label{cri01}
e^{-\beta\Delta(\sigma,\eta)-\beta\gamma(\beta)}
\le
p(\sigma,\eta)
\le
e^{-\beta\Delta(\sigma,\eta)+\beta\gamma(\beta)}
\end{equation}
with $\gamma(\beta)\to0$ in the zero temperature limit $\beta\to\infty$.
Hence, $\Delta$ can be 
interpreted as the cost of the transition from $\sigma$ to $\eta$
and plays the role that, in the context of Glauber dynamics, is 
played by the difference of energy. In this context 
the ground states are those configurations on which the Gibbs
measure $\mu$ concentrates when
$\beta\to\infty$; hence, they can be defined as the
minima of the \textit{energy}:
\begin{equation}
\label{hl}
E(\sigma)=
\lim_{\beta\to\infty}H(\sigma)
=
-h\sum_{x\in\Lambda}\sigma(x)
-\sum_{x\in\Lambda}|S_{\sigma}(x)+h|
\end{equation}
For $h>0$ the configuration $\puno$, with
$\puno(x)=+1$ for $x\in\Lambda$, is the unique ground state,
indeed each site contributes to the energy with $-h-(4+h)$. 
For $h=0$, the configuration $\muno$, with
$\muno(x)=-1$ for $x\in\Lambda$, is a ground state as well, as all the other configurations such that all the sites
contribute to the sum (\ref{hl}) with $4$. 
Hence,  the checkerboard configurations $\eC,\oC\in\mathcal{S}$ such that 
$\eC(x)=(-1)^{x_1+x_2}$ and $\oC(x)=(-1)^{x_1+x_2+1}$
for $x=(x_1,x_2)\in\Lambda$ are ground states, as well. Notice that 
$\eC$ and $\oC$ are checkerboard--like states with the pluses  
on the even and odd sub--lattices, respectively; we set $\cC=\{\eC,\oC\}$. 
Since the side length $L$ of the torus $\Lambda$ 
is even, then $E(\eC)=E(\oC)=E(\cC)$ (we stress the abuse 
of notation $E(\cC)$).
Under periodic boundary conditions,
we get for the energies:
$E(\puno)=-L^2(4+2h)$, 
$E(\muno)=-L^2(4-2h)$, 
and
$E(\cC)=-4L^2$. Therefore,
\begin{equation}
\label{energy_dis}
E(\muno)>E(\cC)>E(\puno)
\end{equation}
for $ 0<h\le1$.
Moreover, using the analysis in \cite{CN} we can derive Fig.~\ref{fig:fig01}, with the series of the two local minima $\muno,\cC$.

\begin{figure}[t]
\centering
\hspace{-6.0cm}
 \scalebox{.95}{
 \begin{picture}(90,110)(40,25)
 \thinlines
 \setlength{\unitlength}{0.08cm}
 \qbezier(50,30)(60,60)(70,60)
 \qbezier(70,60)(80,60)(85,20)
 \put(70,50){\vector(0,1){10}}
 \put(70,40){\vector(0,-1){10}}
 \put(68,43){${\Gamma}$}
 \put(48,24){${\muno}$}
 \put(84,14){${\cC}$}
 \qbezier(85,20)(90,50)(95,50)
 \qbezier(95,50)(100,50)(105,15)
 \put(95,40){\vector(0,1){10}}
 \put(95,30){\vector(0,-1){10}}
 \put(93,33){${\Gamma}$}
 \put(104,10){${\puno}$}
 \end{picture}}
 \caption{Schematic description of the energy landscape for a series of 
          metastable states. Note that the ground state is $\puno$ and 
$E(\muno)>E(\cC)>E(\puno)$.}
 \label{fig:fig01}
\end{figure}


We conclude this section by listing some relevant definitions.
Given $\sigma\in\cc{S}$ we consider the chain with
initial configuration $\sigma_0=\sigma$, we denote with
$\bb{P}_\sigma$ the probability measure on the space
of trajectories, by $\bb{E}_\sigma$ the corresponding expectation value,
and by
\begin{equation}
\label{hitting}
\tau_A^\sigma:=\inf\{t>0:\,\sigma_t\in A\}
\end{equation}
the {\it first hitting time on} $A\subset\cc{S}$; we shall drop the initial
configuration from the notation (\ref{hitting}) whenever it is equal to
$\muno$, we shall write $\tau_A$ for $\tau_A^{\muno}$, namely.
Moreover, a finite sequence of configurations
$\omega=\{\omega_1,\dots,\omega_n\}$ is called the
{\it path} with starting configuration $\omega_1$ and ending
configuration $\omega_n$; we let $|\omega|:=n$. 
Given a path $\omega$ we define the {\it height} along $\omega$ as:
\begin{equation}
\label{pizza}
\Phi_{\omega}:=H(\omega_1) \;\textrm{ if } |\omega|=1
\;\;\textrm{ and }\;\;
\Phi_{\omega}:=\max_{i=1,\dots,|\omega|-1} H(\omega_i,\omega_{i+1})
\;\;\textrm{ otherwise}
\end{equation}
where $H(\omega_i,\omega_{i+1})$ is the \emph{communication height} between the configurations 
$\omega_i$ and $\omega_{i+1}$, defined as follows:
\begin{equation}
\label{heights}
H(\omega_i,\omega_{i+1}):= H(\omega_i)-\frac{1}{\beta}\log(p(\omega_i,\omega_{i+1}))
\end{equation} 
Given two configurations $\sigma,\eta\in\cc{S}$, we denote by
$\Theta(\sigma,\eta)$ the set of all the paths $\omega$ starting from
$\sigma$ and ending in $\eta$. The minimax between $\sigma$ and $\eta$ is
defined as
\begin{equation}
\label{pizza2}
\Phi(\sigma,\eta):=
\min_{\omega\in\Theta(\sigma,\eta)}\Phi_\omega
\end{equation}

\section{Metastable states and main results} 
\label{meta}
\begin{figure}[t]
\centering
\label{fig:landscape}
\vskip 0.5cm
\includegraphics[width=7cm,height=3.5cm]{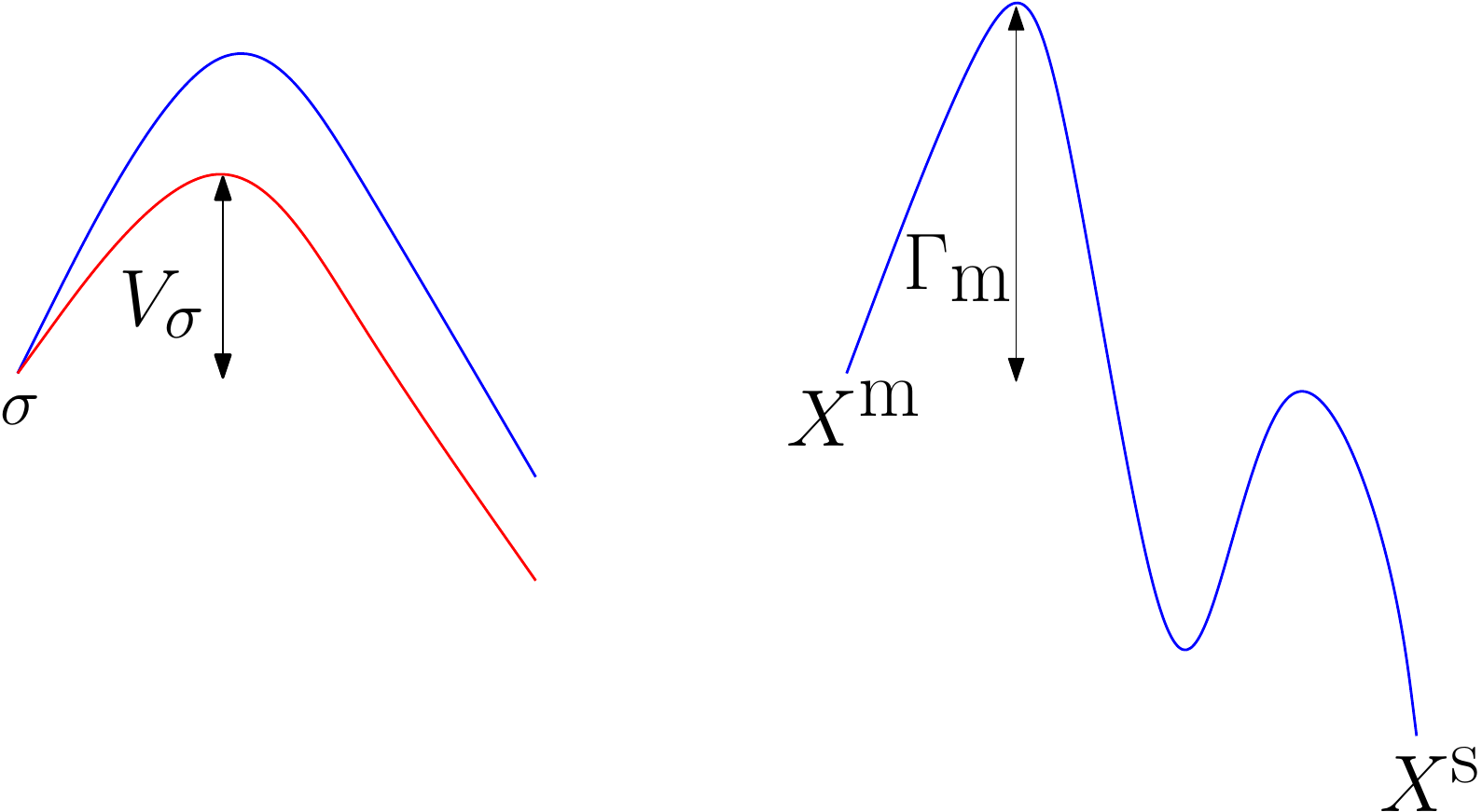}
\caption{Definition of metastable states.}
\label{fig:definizione}
\end{figure}

\\
We want now to define the notion of metastable states. 
See Fig.~\ref{fig:definizione} for a graphic description 
of the quantities we are going to define.  
For any $\sigma\in\mathcal{S}$, we let
$\mathcal{I}_\sigma\subset\mathcal{S}$
be the set of configurations with energy strictly below $H(\sigma)$ and
$V_\sigma=\Phi(\sigma,\mathcal{I}_\sigma)-H(\sigma)$ 
be the \textit{stability level
of} $\sigma$, that is the energy barrier that, starting from $\sigma$, must be 
overcome to reach the set of configurations with energy smaller than 
$H(\sigma)$.
We denote by $X^\textrm{s}$ the set of
global minima of the energy, i.e., the collection of the 
ground states, and suppose that 
the \textit{communication energy}
$\Gamma=\max_{\sigma\in\mathcal{S}\setminus X^\textrm{s}}V_\sigma$
is strictly positive.
Finally, we define the set of \textit{metastable states}
$X^\textrm{m}=\{\eta\in\mathcal{S}:\,V_\eta=\Gamma\}$.
The set $X^\textrm{m}$ deserves its name, 
since in a rather general framework it is possible to prove 
(see, e.g., \cite[Theorem~4.9]{MNOS}) the following:
pick $\sigma\in X^\textrm{m}$, consider the 
chain $\sigma_n$ started at 
$\sigma_0=\sigma$, then the \textit{first hitting time}
$\tau_{X^\textrm{s}}$ 
to the ground states is a random variable with mean 
exponentially large in $\beta$, that is 
\begin{equation}
\label{mnos4.9}
\lim_{\beta\to\infty}
\frac{1}{\beta}\,\log\mathbb{E}_\sigma[\tau_{X^\textrm{s}}]=\Gamma
\end{equation}
In the considered regime, finite volume and temperature tending to zero, 
the description of metastability is then reduced
to the computation of $X^\textrm{s}$, $\Gamma$, and 
$X^\textrm{m}$.

We pose now the problem of metastability and state the related theorem
on the sharp estimates for the exit time.
Consider the model (\ref{markov}) with $0<h<1$ and
suppose that the system is prepared in the state $\sigma_0=\muno$, 
and we estimate the first time at which the system reaches $\puno$. As showed in \cite{CN}, the system visits
with probability tending to one in the $\beta\to\infty$ limit 
the checkerboards $\cC$, and the typical time to jump from $\muno$ to $\cC$ is the
same as the time needed to jump from $\cC$ to $\puno$. Hence, the aim of this paper is to prove an addition formula for 
the expected exit times from $\muno$ to $\puno$. The metastable states $\muno$ and $\cC$ 
form indeed a \emph{series}: the system started at $\muno$ must necessarily pass through $\cC$ before relaxing to the 
stable state $\puno$.

In order to state the main theorem, we have to introduce the
following \emph{activation energy} $\Gamma_\textrm{m}$, which corresponds to 
the energy of the critical configuration triggering the nucleation: 
\begin{eqnarray}
\Gamma_\textrm{m}&:=&
-2h\lambda^2+2(4+h)\lambda-2h \label{prot-h2}
\end{eqnarray}
where $\lambda$  is the {\it critical length} defined as 
$\lambda:=\lfloor2/h\rfloor+1$.
In other words, in \cite{CN} it is proven that with probability 
tending to one in the limit $\beta\to\infty$, before reaching 
the checkerboards $\cC$, the system necessarily visits a particular 
set of configurations, called \emph{critical droplets}, which are 
all the configurations (equivalent up to translations, rotations, 
and protuberance possible shifts) 
with a single checkerboard droplet in the sea of minuses with 
the following shape: a rectangle of side lengths $\lambda$ and 
$\lambda-1$ with a unit square attached to one of the two longest 
sides (the protuberance) and with the spin in the protuberance being 
plus. This way of escaping from the metastable 
via the formation of a single droplet is called \emph{nucleation}.

In order to study the exit times from the from one of the two metastable states towards the stable configuration,
we use a decomposition valid for a general Markov chains derived 
in \cite{CNS04} and that we recall for the sake of completeness.
\begin{lemma}
\label{t:dimo00}
Consider a finite state 
space $X$ and a family of irreducible and aperiodic Markov chain. 
Given three states $y,w,z\in X$ pairwise mutually different, we have
that the following holds
\begin{equation}
\label{dimo00}
\bb{E}_y[\tau_z]
=
\bb{E}_y[\tau_w\mathbf{1}_{\{\tau_w<\tau_z\}}]
+
\bb{E}_w[\tau_z]\bb{P}_y(\tau_w<\tau_z)
+
\bb{E}_y[\tau_z\mathbf{1}_{\{\tau_w\geq\tau_z\}}]
\end{equation}
\end{lemma}
where $\bb{E}_x[\cdot]$ denotes the average along the trajectories
of the Markov chain started at $x$, and $\tau_y$ is the first hitting 
time to $y$ for the chain started at $x$.
In the expressions above $\mathbf{1}_{\{\cdot\}}$ is the \emph{characteristic}
function which is equal to one if the event $\{\cdot\}$ is realized and 
zero otherwise. 

\begin{theorem}\label{sharp2}
Consider the PCA (\ref{markov}), for
$h>0$ small enough, and $\beta$ large enough,  we have:
\begin{equation}
\label{t:main}
\bb{E}_{\muno}(\tau_{\puno})=\left(\frac{1}{k_1}+\frac{1}{k_2}\right) 
e^{\beta\Gamma_{\normalfont\textrm{m}}}[1+o(1)],
\end{equation}
where $o(1)$ denotes a function going to zero when $\beta\to\infty$ and
$$
k_1=k_2=K=8\lambda \vert\Lambda|
$$
\end{theorem}

The term 
$e^{\beta\Gamma_\textrm{m}}/k_1$  in  (\ref{t:main})
represents the contribution of the mean hitting time 
 $\bb{E}_\muno[\tau_\cC \mathbf{1}_{\{\tau_\cC<\tau_\puno\}}]
$  to the relation (\ref{dimo00}) 
and $e^{\beta\Gamma_{\normalfont\textrm{m}}}/k_2$ the contribution of $\bb{E}_{\cC}\tau_{\puno}$. The pre--factors $k_1$ and $k_2$ give the precise estimate of the mean nucleation time of the stable phase,  beyond the asymptotic exponential regime and represent entropic factors, counting the cardinality of the critical droplets which trigger the nucleation. 
At the level of logarithmic equivalence, namely, by renouncing 
to get sharp estimate, this result can be proven by 
the methods in \cite{MNOS}. More precisely, one gets that 
$(1/\beta)\log\bb{E}_\muno[\tau_\puno]$ tends to $\Gamma_\rr{m}$ 
in the large $\beta$ limit.
\subsection{Potential theoretic approach and capacities}
\par\noindent
Since our results on the precise asymptotic of the mean nucleation time of the stable phase are strictly related to the potential theoretic approach to metastability (see \cite{BEGK}), we recall some definitions and notions.
We define the \textit{Dirichlet form}  
associated to the reversible Markov chain, with transition probabilities $p(\sigma,\eta)$ and equilibrium measure $\mu$, as the functional:
\begin{equation}\label{diri}
\cc{E}(h)=\frac{1}{2}\sum_{\sigma,\eta\in\cc{S}} \mu(\sigma) p(\sigma, \eta) {[f(\sigma)-f(\eta)]}^2,
\end{equation}

\noindent where $f:\cc{S}\to[0,1] $ is a generic function.
The form (\ref{diri}) can be rewritten in terms of the communication heights $H(\sigma,\eta)$ and of the partition function $Z$:

\begin{equation}
 \cc{E}(h)=\frac{1}{2}\sum_{\sigma,\eta\in\cc{S}} \frac{1}{Z}\, e^{-\beta H(\sigma,\eta)} {[f(\sigma)-f(\eta)]}^2.
\end{equation}
\noindent Given two non-empty disjoint sets $\cc{A}, \cc{B}$ the \textit{capacity} of the pair $\cc{A}, \cc{B}$ is defined by
\begin{equation}
\label{def_cap}
   \rr{cap}_\beta(\cc{A},\cc{B}):= \min_{{f:\cc{S}\to[0,1]} \atop{f\vert_\cc{A}=1,f\vert_\cc{B}=0}} \cc{E}(f)  
\end{equation}
\noindent and from this definition it follows that the capacity is a \textit{symmetric} function of the sets $\cc{A}$ and $\cc{B}$.\\
The right hand side of (\ref{def_cap}) has a unique minimizer $f_{\cc{A},\cc{B}}^\star$ called \textit{equilibrium potential} of the pair $\cc{A}, \cc{B}$ given  by
\begin{equation}
f_{\cc{A},\cc{B}}^\star(\eta)=\bb{P}_\eta(\tau_\cc{A}<\tau_\cc{B}),
\end{equation}
for any $\eta\notin \cc{A}\cup\cc{B}$.
The strength of this variational representation comes from the \emph{monotonicity} of the Dirichlet form
in the variable $p(\sigma,\eta)$. In fact, the Dirichlet form $\cc{E}(f)$ is a monotone increasing function of the transition probabilities $p(x,y)$ for $x\neq y$, while it is independent on the value $p(x,x)$.
In fact, the following theorem holds:
\begin{theorem} 
\label{mono}
Assume that $\cc{E}$ and $\tilde{\cc{E}}$ are Dirichlet forms 
associated to two Markov chains
$P$ and $\tilde{P}$ with state space $\cc{S}$ and reversible 
with respect to the measure $\mu$. Assume that the transition 
probabilities $p$ and $\tilde{p}$ are
given, for $x\neq y$, by
\begin{equation*}
p(x,y)=g(x,y)/\mu(x)
\,\,\,\textrm{ and }\,\,\,
\tilde{p}(x,y)=\tilde{g}(x,y)/\mu(x)
\end{equation*}
where $g(x,y)=g(y,x)$ and 
$\tilde{g}(x,y)=\tilde{g}(y,x)$, and, for all $x\neq y$,
$\tilde{g}(x,y)\leq g(x,y)$.
Then, for any disjoint sets $\mathcal{A},\,\mathcal{D}\subset\cc{S}$ we have:
\begin{equation}
\rr{cap}_\beta(\mathcal{A},\mathcal{D})\geq \widetilde{\rr{cap}_\beta}(\mathcal{A},\mathcal{D})
\end{equation}
\end{theorem}
We will use Theorem \ref{mono} by simply setting some of the transition probabilities $p(x,y)$ equal to zero.
Indeed if enough of these are zero, we obtain a chain where everything can be computed easily. In order to get
a good lower bound, the trick will be to guess which transitions can be switched off without altering
the capacities too much, and still to simplify enough to be able to compute it.

\subsection{Series of metastable states for PCA without self--interaction}
\label{s:series}
\par\noindent
In this section we  state the model--dependent results for the class of PCA  considered, which, by the general theory contained in \cite{CNS04},  imply Theorem~\ref{sharp2}.
\begin{lemma}
\label{l:series00}
The configurations
 $\muno$, $\cC$, and $\puno$ are 
such that 
$X^\rr{s}=\{\puno\}$,
$X^\rr{m}=\{\muno,\cC\}$,
$E(\muno)>E(\cC)$, and $\Gamma=\Gamma_{\textnormal{m}}$.
\end{lemma}
Our model presents 
the series structure depicted in Fig.~\ref{fig:fig01}:
when the system is started at $\muno$ with high 
probability it will visit $\cC$ before $\puno$. 
In fact, the following lemma holds:

\begin{lemma}
\label{l:series01}
There exists $\lambda>0$ and $\beta_0>0$ such that for any $\beta>\beta_0$ 
\begin{equation}
\label{series01}
\bb{P}_{\muno}(\tau_\puno<\tau_\cC)\le e^{-\beta\lambda}
\end{equation}
\end{lemma}
We use Lemma~\ref{t:dimo00} and Lemma~\ref{l:series01} and, 
in order to drop the last addendum of (\ref{t:dimo00}),
we  need  an exponential control of the tail 
of the distribution of the suitably 
rescaled random variable $\tau_\stab$. 
\begin{lemma}
\label{l:series015}
For any $\delta>0$ there exists $\beta_0>0$ such that   
\begin{equation}
\label{series015}
\sum_{t=0}^\infty 
 t\;\bb{P}_\muno(\tau_\puno> t e^{\beta\Gamma_\rr{m}+\beta\delta})
\le 
1/3
\end{equation}
for any $\beta>\beta_0$. 
\end{lemma}

By Lemma~\ref{t:dimo00}, Lemma~\ref{l:series01}, Lemma~\ref{l:series015} we have that:
\begin{equation}
\label{e:finale}
\bb{E}_\muno[\tau_\puno]
=\left(
\bb{E}_\muno[\tau_\cC\mathbf{1}_{\{\tau_\cC<\tau_\puno\}}]
+
\bb{E}_\cC[\tau_\puno] \right) [1+o(1)]
\end{equation}
The next lemma regards the estimation of the two addenda of (\ref{e:finale}), using the potential theoretic approach:
\begin{lemma}
\label{l:series02}
There exist two positive constants $k_1,k_2<\infty$ such that 
\begin{equation}
\label{series02}
\frac{\mu(\muno)}{\rr{cap}_\beta(\muno,\{\cC,\puno\})}
=
\frac{e^{\beta\Gamma_\rr{m}}}{k_2}[1+o(1)]
\;\;\textrm{ and }\;\;
\frac{\mu(\cC)}{\rr{cap}_\beta(\cC,\puno)}
=
\frac{e^{\beta\Gamma_\rr{m}}}{k_1}[1+o(1)]
\;\;.
\end{equation}
\end{lemma}
By general standard results of  the potential theoretic approach, it can be shown indeed that $\bb{E}_\muno[\tau_\cC\mathbf{1}_{\{\tau_\cC<\tau_\puno\}}]$ 
equals the left hand side in the first of (\ref{series02}) 
and $\bb{E}_\cC[\tau_\puno]$ equals 
the left hand side in the second 
of (\ref{series02}). Hence, 
by (\ref{e:finale}), Lemmata~\ref{l:series00}, \ref{l:series01}, 
\ref{l:series015} and \ref{l:series02}, Theorem~\ref{sharp2} follows.


\section{Sketch of proof of Theorem~\ref{sharp2}} 
\label{s:proof}
\par\noindent 
In this section we prove Theorem~\ref{sharp2}, by proving Lemmata~\ref{l:series00}, \ref{l:series01}, 
\ref{l:series015}, 
and \ref{l:series02}.  

Due to space constraints we sketch the main idea behind the proof of Lemmata  \ref{l:series00}, \ref{l:series01}, 
\ref{l:series015} and we give in detail the proof of \ref{l:series02}.
As regards  Lemma~\ref{l:series00}, the energy inequalities and $X^\rr{s}=\{\puno\}$, 
 easily follow by (\ref{energy_dis}). However, in order to prove $X^\rr{m}=\{\muno, \cC\}$, for any $\sigma\in\mathcal{S}\setminus(X^\rr{s}\cup X^\rr{m})$,
 we have to show that there exists a path
  $\omega:\sigma\to\mathcal{I}_\sigma$ such that the maximal \emph{communication height} to overcome to reach a configuration at lower energy is smaller than 
  $\Gamma_\rr{m}+E(\sigma)$, i.e. $\Phi_\omega<\Gamma_\rr{m}+E(\sigma)$.  
By Prop.\ 3.3 of Ref.\ \cite{CN}  
for all configurations $\sigma$ there exists a downhill path to a configuration consisting of union of rectangular droplets: for instance rectangular checkerboard in a see of minuses or well separated plus droplets in a sea of minuses or inside a checkerboard droplet. In case these droplets are non--interacting (i.e. at distance larger than one), by the  analysis of the growth/shrinkage mechanism of rectangular droplets contained in \cite{CN}, it is straightforward to find the required path. In case of interacting rectangular droplets a more accurate analysis is required, but this is outside the scope of the present paper. Lemma~\ref{l:series01} is a consequence of the \emph{exit tube} results contained in \cite{CN}, while Lemma
~\ref{l:series015} follows by Th. 3.1 and (3.7) in \cite{MNOS} with an appropriate constant.

\subsection{Proof of Lemma~\ref{l:series02}}
\label{capaci}
\par\noindent
We recall the definition of the cycle $\cc{A}_{\muno}$ playing the role of a generalized \textit{basin of attraction} of the $\muno$ phase:
\begin{eqnarray}
\cc{A}_{\muno}
&:=&
\{ \eta\in\cc{G}_{\muno}:\exists \omega=\{\omega_0=\eta,...,\omega_n=\muno \}
\textnormal{ such that }\, 
 \omega_0,...,\omega_n\in  \cc{G}_{\muno}  \nonumber
\\
&\!\!\!&\!\!\!
\phantom{\{m}
\textnormal{and }\, \Phi_\omega<\Gamma +E(\muno) \}
\nonumber
\end{eqnarray}
where $\cc{G}_{\muno}$ is the set defined in [Sect. 4,\cite{CN}], containing the \emph{sub--critical} configurations (e.g. a single checkerboard rectangle in a see of minuses, with shortest side smaller than the critical length $\lambda$). In a very similar way, we can define
\begin{eqnarray*}
\cc{A}_{\{\puno,\cC\}}
&:=&
\{ \eta\in\cc{G}_{\muno}^c:\exists 
\omega=\{\omega_0=\eta,...,\omega_n\in\{\cC,\puno\}\}
\textnormal{ such that }\, 
\\
&\!\!\!&\!\!\!
\phantom{\{m}
 \omega_0,...,\omega_n\in  \cc{G}_{\muno}^c  \nonumber
\textnormal{and }\, \Phi_\omega<\Gamma +E(\muno) \}
\nonumber
\end{eqnarray*}
We start proving the equality on the left in 
(\ref{series02}) by giving an upper and lower estimate for the capacity $\textnormal{cap}_\beta(\muno,\{\puno,\cC\})$. 
\noindent Thus, what we need is the precise estimates on capacities, via sharp upper and lower bounds.

Usually the upper bound is the simplest because  it can be given by choosing a suitable  test function. Instead, for the lower bound, we use the monotonicity of the Dirichlet form in the transition probabilities via simplified processes.
Therefore, we firstly identify the domain where $f^\star$ is close to one and to zero, in our case the set $\cc{A}_{\muno}$ and $\cc{A}_{\{\puno,\cC\}}$ respectively. Restricting the processes on these sets 
and by rough estimates on capacities we are able to give a sharper lower 
bound for the capacities themselves.

%
%
%
\underline{Upper bound.}
We use the general strategy to prove an upper bound by guessing some 
a priori properties
of the minimizer, $f^\star$, and then to find the minimizers within this class.
Let us consider the two basins of attraction $\cc{A}_{\muno}$ and $\cc{A}_{\muno,\{\cC,\puno\}}$. A potential $f^u$ will provide an upper bound for the capacity, i.e. the Dirichlet form evaluated at the equilibrium
potential $f^\star_{\muno,\{\cC,\puno\}}$, solution of the variational problem (\ref{def_cap}), where the two sets with the boundary conditions are $\muno$ and 
$\{\cC,\puno\}$. We choose the following test function for giving an upper bound for the capacity:
\begin{equation}
f^u(x):=\left\{
\begin{tabular}{c c}
$1$\,\, & $x\in\cc{A}_{\muno}$,\\
$0$\,\,& $x\in \cc{A}_{\muno}^c$
\end{tabular}
\right.
\end{equation}
so that:
\begin{equation}
\label{upper_b}
\cc{E}(f^u)=\frac{1}{Z}
\sum_{{\sigma\in \cc{A}_{\muno},}\atop{ \eta\in \cc{A}_{\{\cC,\puno\}}}} 
\!\!\! 
e^{-\beta H(\sigma,\eta)}
+\frac{1}{Z}\sum_{{\sigma\in \cc{A}_{\muno},}\atop{ \eta\in {(\cc{A}_{\{\cC,\puno\}}\cup\cc{A}_{\muno})}^c}} 
\!\!\!\!\!\! 
e^{-\beta H(\sigma,\eta)}
\end{equation}
Therefore, by Lemma 4.1 in \cite{CN}, (\ref{upper_b}) can be easily bounded by:
\begin{equation}
\cc{E}(h^u)/\mu(\muno)\leq K\, e^{-\beta\Gamma}+\vert\cc{S}\vert\,e^{-\beta(\Gamma+\delta)},
\end{equation}
where $\delta>0$, because  $\Gamma+E(\muno)$ is nothing but the minmax between $\cc{A}_{\muno}$ and its complement and $E(\sigma,\eta)=\Gamma+E(\muno)$ only in the transition between configurations belonging to $\cc{P}^\prime\subset \cc{A}_{\muno}$, and some particular configurations belonging to $\cc{P}\subset\cc{A}_{\{\cC,\puno\}}$ for all the other transitions we have $E(\sigma,\eta)>\Gamma+E(\muno)$. In particular, $\cc{P}^\prime$ is the set of configurations consisting of rectangular checkerboard $R_{\lambda,\lambda-1}$ of sides $\lambda$ and $\lambda-1$ in a see of minuses.  $\cc{P}$ is instead the subset of \emph{critical configurations} obtained  by  flipping a single site 
adjacent to a plus spin of the internal checkerboard along the larger side of a configuration $\eta\in \cc{P}^\prime$.

\underline{Lower bound.}
In order to have a lower bound, let us estimate the equilibrium potential.
We can prove the following Lemma:
\begin{lemma} \label{anton2}
$\exists C,\delta>0$ such that for all $\beta>0$
\begin{equation}
\label{anton2.2}
\min_{\eta\in\cc{A}_{\muno}}f^\star(\eta)\geq 1-Ce^{-\delta\beta} 
\;\;\textrm{ and }\;\;
\max_{\eta\in\cc{A}_{\{\cC,\puno\}}}f^\star(\eta)\leq Ce^{-\delta\beta}
\end{equation}
\end{lemma}
\noindent \textit{Proof.\/}
Using a standard renewal argument, given $\eta\notin\{\muno,\cC,\puno\}$:
\begin{displaymath}
\bb{P}_\eta(\tau_{\{\cC,\puno\}}<\tau_{\muno})
=
\frac{\bb{P}_\eta(\tau_{\{\cC,\puno\}}<\tau_{\muno\cup\eta})}{1-\bb{P}_\eta(\tau_{\muno\cup{\{\cC,\puno\}}}>\tau_{\eta})}
\end{displaymath}
and
\begin{displaymath}
\bb{P}_\eta(\tau_{\muno}<\tau_{\{\cC,\puno\}})
=\frac{\bb{P}_\eta(\tau_{\muno}<\tau_{{\{\cC,\puno\}}\cup\eta})}{1-\bb{P}_\eta(\tau_{\muno\cup{\{\cC,\puno\}}}>\tau_{\eta})}.
\end{displaymath}
If the process started at point $\eta$ wants to realize indeed the event $\{\tau_{\{\cC,\puno\}}<\tau_{\muno}\}$ it can  either go to $\muno$ immediately and without returning to $\eta$ again, or it may return to $\eta$ without going to ${\{\cC,\puno\}}$ or $\muno$. Clearly, once the process returns to $\eta$, we can use the strong Markov property. Thus
\begin{eqnarray*}
\bb{P}_\eta(\tau_{\{\cC,\puno\}}<\tau_{\muno})
&=&
\bb{P}_\eta(\tau_{\{\cC,\puno\}}<\tau_{\muno\cup\eta})+\bb{P}_\eta(\tau_\eta<\tau_{{\{\cC,\puno\}}\cup\muno}\wedge\tau_{\{\cC,\puno\}}<\tau_{\muno}) \\
&=& 
\bb{P}_\eta(\tau_{\{\cC,\puno\}}<\tau_{\muno\cup\eta})\\
&&+\bb{P}_\eta(\tau_\eta<\tau_{{\{\cC,\puno\}}\cup\muno})
\bb{P}_\eta(\tau_{\{\cC,\puno\}}<\tau_{\muno}) 
\end{eqnarray*}
and, solving the equation for  $\bb{P}_\eta(\tau_{\{\cC,\puno\}}<\tau_{\muno})$, we have the renewal equation.
Then $\forall \eta\in\cc{A}_{\muno}\setminus\{\muno\}$ we have:
\begin{displaymath}
f^\star(\eta)=1-\bb{P}_\eta(\tau_{\{\cC,\puno\}}<\tau_{\muno}\})=1-\frac{\bb{P}_\eta(\tau_{\{\cC,\puno\}}<\tau_{\muno\cup\eta})}{\bb{P}_\eta(\tau_{\muno\cup{\{\cC,\puno\}}}<\tau_{\eta})}
\end{displaymath}
and, hence, 
\begin{displaymath}
f^\star(\eta)
\geq 1-\frac{\bb{P}_\eta(\tau_{\{\cC,\puno\}}<\tau_{\eta})}{\bb{P}_\eta(\tau_{\muno}<\tau_{\eta})} 
\end{displaymath}
For the last term we have the equality:
\begin{equation}\label{anton1}
\frac{\bb{P}_\eta(\tau_{\{\cC,\puno\}}<\tau_{\eta})}
{\bb{P}_\eta(\tau_{\muno}<\tau_{\eta})}=\frac{\textnormal{cap}_\beta(\eta,{\{\cC,\puno\}})}{\textnormal{cap}_\beta(\eta,\muno)}
\end{equation}
The upper bound for the numerator of ($\ref{anton1}$) is easily obtained through the upper bound on
$\textnormal{cap}_\beta(\muno,{\{\cC,\puno\}})$ which we already have. The lower bound on the denominator is obtained by reducing the state space to a single path  from $\eta$ to $\muno$, picking an \emph{optimal} path $\omega=\{\omega_0,\omega_1,...,\omega_N\}$ that realizes the minmax $\Phi(\eta,\muno)$ and ignoring all the
transitions that are not in the path. Indeed  by Th.~ \ref{mono}, we use the monotonicity of the Dirichlet form
in the transition probabilities $p(\sigma,\eta)$, for $\sigma\neq \eta$. Thus, we can have a lower bound for capacities by simply setting
some of the transition probabilities $p(\sigma,\eta)$  equal to zero. It is clear that if enough of these are set to zero, we obtain a chain where everything can be computed easily. With our choice we have:
\begin{equation}
\textnormal{cap}_\beta(\eta,\muno)\geq \min_{{f:\omega\to [0,1]}\atop{f(\omega_0)=1, f(\omega_N)=0}}\cc{E}^\omega(f)
\end{equation}

\noindent where the Dirichlet form $\cc{E}^\omega(f)$ is defined as $\cc{E}$ in  (\ref{diri}), with $\cc{S}$ replaced by $\omega$. Due to the one--dimensional nature of the set $\omega$, the variational problem in the right hand side can be solved explicitly by elementary computations. One finds 
that the minimum equals
 \begin{equation}
 M={\left[\sum_{k=0}^{N-1}Z\,e^{\beta H(\omega_k,\omega_{k+1})}\right]}^{-1}
  \end{equation}
and it is uniquely attained at $f$ given by
\begin{equation}
f(\omega_k)=M\sum_{l=0}^{k-1}Z\,e^{\beta H(\omega_l,\omega_{l+1})}\,\,\,\,\,\,\,\,\,\,\,\,\,k=0,1,...,N.
\end{equation}
Therefore,
\begin{eqnarray*}
\textnormal{cap}_\beta(\eta,\muno) &\geq& M
\geq\frac{1}{K\,Z}\max_{k} e^{-\beta H(\omega_k,\omega_{k+1})}
\end{eqnarray*}
and hence
\begin{displaymath}
\textnormal{cap}_\beta(\eta,\muno) \geq 
C_1\,\frac{1}{Z} e^{-\beta\Phi(\eta,\muno)}
\end{displaymath}
with $\lim_{\beta\to\infty} C_1=1/K$.
Moreover, 
we know that if $\eta\in \cc{A}_{\muno}$ then it holds 
$\Phi(\eta,\muno)<\Phi(\eta,\{\cC,\puno\})$. Indeed, by the definition of the set $\cc{A}_{\muno}$:
\begin{equation}
\Phi(\eta,\{\cC,\puno\})\geq \Gamma+E(\muno) > \Phi(\eta,\muno).
\end{equation}
For this reason
\begin{displaymath}
f^\star(\eta)\geq 1-C(\eta) e^{-\beta(\Phi(\eta,\{\cC,\puno\})-\Phi(\eta,\muno))}\geq 1-C(\eta)e^{-\beta\delta},
\end{displaymath}
and we can take  $C:=\sup_{\eta\in\cc{A}_{\muno}\setminus\muno} C(\eta)$. 
Otherwise $\forall \eta\in\cc{A}_{\{\cC,\puno\}}\setminus\{\cC,\puno\}$ we have:
\begin{displaymath}
f^\star(\eta)
=\bb{P}_\eta(\tau_{\muno}<\tau_{\{\cC,\puno\}}\})
=\frac{\bb{P}_\eta(\tau_{\muno}<\tau_{\{\cC,\puno\}\cup\eta})}
{\bb{P}_\eta(\tau_{\muno\cup\{\cC,\puno\}}<\tau_{\eta})}
\end{displaymath}
and hence
\begin{displaymath}
f^\star(\eta)
\leq\frac{\bb{P}_\eta(\tau_{\muno}<\tau_{\eta})}{\bb{P}_\eta(\tau_{\{\cC,\puno\}}<\tau_{\eta})}\nonumber 
= \frac{\textnormal{cap}_\beta(\eta,\muno)}{\textnormal{cap}_\beta(\eta,\{\cC,\puno\})}\leq C(\eta) e^{-\beta\delta}
\end{displaymath}
\noindent 
proving the second equality \eqref{anton2.2} with 
$C=\max_{\eta\in\cc{A}_{\{\cC,\puno\}}\setminus\{\cC,\puno\}} C(\eta)$.
\hfill 
\qed

\begin{figure}[t]
\centering
\label{fig:saddles}
\vskip 0.5cm
\includegraphics[height=3.8cm]{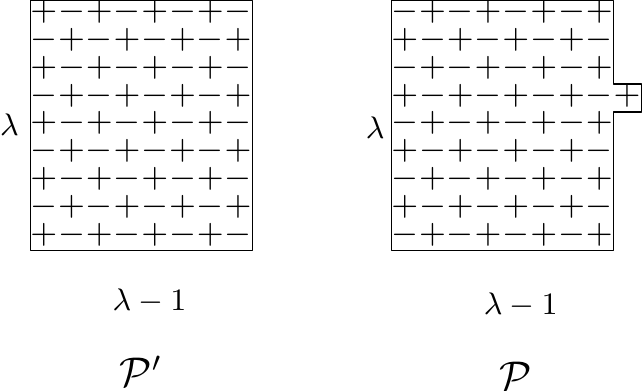}
\caption{\emph{Saddles} configurations.}
\label{fig:definizione2}
\end{figure}

Now we are able to give a lower bound for the capacity.
By (\ref{def_cap}), we have:
\begin{eqnarray*}
\textnormal{cap}_\beta(\muno,\{\cC,\puno\})&=&\cc{E}(f^\star)
\geq\sum_{{\sigma\in\mathcal{A}_{\muno}}\atop{\eta\in\mathcal{A}_{\{\cC,\puno\}}}}\mu(\sigma) p(\sigma,\eta){(f^\star(\sigma)-f^\star(\eta))}^2 \\
&\geq& K\mu(\muno)e^{-\beta\Gamma}+o(e^{-\beta\delta})
\end{eqnarray*}
\noindent Now we want to evaluate the combinatorial pre--factor $K$ of the sharp estimate. 
We have to determinate all the possible ways to choose a critical droplet in the lattice with periodic boundary conditions. We know that the set $\cc{P}$ of such configurations contains all the checkerboard rectangles $R_{\lambda-1,\lambda}$ in a see of minuses (see Fig.~\ref{fig:definizione2}). 
Because of the translational invariance on the lattice, we can associate at each site $x$ two rectangular droplets $R_{\lambda-1,\lambda}$ and $R_{\lambda,\lambda-1}$ such that their north-west corner is in $x$. Considering the periodic boundary conditions and being $\Lambda$ a square of side $L$, 
the number of such rectangles is $A=2L^2$
In order to calculate $K$, we have to count in how many ways  we can add a protuberance to a rectangular  checkerboard configuration  $R_{\lambda-1,\lambda}$, along the largest side and  adjacent to a plus spin of the checkerboard. Hence, we have that 
$K=4A\lambda =8\lambda L^2$, 
and this completes the proof  of the first equality in (\ref{series02}). 
The proof of the second equality in (\ref{series02}) 
can be achieved using very similar arguments.
\hfill \qed

\vspace{0.3cm}
\par\noindent
\textbf{Acknowledgements.\/}
The authors thank A.\ Bovier, F.\ den Hollander, M.\ Slowick, and 
A.\ Gaudilli\`ere
for valuable discussions.

\end{document}